# How a Single Paper Affects the Impact Factor: Implications for Scholarly Publishing

Manolis Antonoyiannakis[1,2]

[1] ma2529@columbia.edu, Department of Applied Physics & Applied Mathematics, Columbia University, 500 W. 120th St., Mudd 200, New York, NY 10027 (USA)

[2] American Physical Society, Editorial Office, 1 Research Road, Ridge, NY (USA)

**Abstract**

Because the Impact Factor (IF) is an average quantity *and* most journals are small, IFs are volatile. We study how a single paper affects the IF using data from 11639 journals in the 2017 Journal Citation Reports. We define as volatility the IF gain (or loss) caused by a single paper, and this is inversely proportional to journal size. We find high volatilities for hundreds of journals annually due to their top-cited paper—whether it is a highly-cited paper in a small journal, or a moderately (or even low) cited paper in a small and low-cited journal. For example, 1218 journals had their most cited paper boost their IF by more than 20%, while for 231 journals the boost exceeded 50%. We find that small journals are rewarded much more than large journals for publishing a highly-cited paper, and are also penalized more for publishing a low-cited paper, especially if they have a high IF. This produces a strong incentive for prestigious, high-IF journals to stay small, to remain competitive in IF rankings. We discuss the implications for breakthrough papers to appear in prestigious journals. We also question the practice of ranking journals by IF given this uneven reward mechanism.

Keywords:

Science of Science – Impact Factor – Volatility – Indicators – Scientific Impact – Citation Distributions

**Introduction**

For a performance indicator of a population of papers to be reliable, it needs to be relatively stable and not highly sensitive to fluctuations or outliers—otherwise, the indicator becomes more of a measure of the few outliers than the general population. So, how volatile are Impact Factors, and other citation averages in general? A single research article can tip the balance in university rankings when citation averages are used (Waltman *et al*., 2011; Bornmann and Marx, 2013), due to the skewed nature of citation distributions. It is also known that in extreme situations, a single paper can strongly boost a journal's IF (Dimitrov, Kaveri, and Bayry, 2010; Moed *et al*., 2012). More recently, Liu *et al*. (2018) studied the effect of a highly-cited paper on the IF of four different-sized journals in particle physics and found that "the IFs of low IF and small-sized journals can be boosted greatly from both the absolute and relative perspectives."

The effect of size of a journal or a university department on its citation average cannot be overstated. Previously (Antonoyiannakis, 2018), we discussed the *overall* influence of journal size on IFs, in the context of the Central Limit Theorem. The Theorem tips the balance in IF rankings, because only small journals can score high IFs, while the IFs of large journals asymptotically approach the global citation average in their field via regression to the mean.

In this paper, first, we introduce the IF volatility index as the change, $\Delta f(c)$—or relative change, $\Delta f_r(c)$—when a *single* paper cited $c$ times is published by a journal of Impact Factor $f$ and size $N$. We study theoretically how $\Delta f(c)$ depends on $c$, $f$, and $N$, and discuss the implications for editorial decisions from the perspective of improving a journal's position in IF rankings. Then, we analyze data from the 11639 journals in the 2017 Journal Citation Reports (JCR) of Clarivate Analytics. We provide summary statistics for the journals' IF volatility to their own top-cited paper. Overall, large values of IF volatility occur for small journal sizes, especially for journals publishing annually fewer than 250 articles or reviews. We discuss the implications for publishing breakthrough papers in high-profile journals.

**How a single paper affects the IF: The general case. Introducing the IF volatility index.**

Here, we consider what happens when a paper that "brings" $c$ citations is published in a journal. The initial IF of the journal is

$$f_1 = \frac{C_1}{N_1}, \qquad (1)$$

where $C_1$ is the number of citations received in a year and $N_1$ is the biennial publication count, i.e., the number of published citable items in the previous 2-year period. With the new paper published by the journal, the new IF becomes

$$f_2 = \frac{C_1 + c}{N_1 + 1}. \qquad (2)$$

The change (volatility) in the IF induced by this one paper is then

$$\Delta f(c) = f_2 - f_1 = \frac{C_1 + c}{N_1 + 1} - \frac{C_1}{N_1} = \frac{c - f_1}{N_1 + 1} \approx \frac{c - f_1}{N_1}, \qquad (3)$$

where the approximation is justified for $N_1 \gg 1$, which applies for all but a few journals that publish only a few items per year. So, the IF change $\Delta f(c)$ depends both on the new paper (i.e., on c) and on the journal (size $N_1$, and citation average $f_1$) where it is published.

We can also consider the *relative change* in the citation average caused by a single paper, which is arguably a more pertinent measure of volatility. That is,

$$\Delta f_r(c) = \frac{f_2 - f_1}{f_1} = \frac{c - f_1}{f_1(N_1 + 1)} \approx \frac{c - f_1}{C_1}, \qquad (4)$$

where, again, the approximation is justified for $N_1 \gg 1$. The above equation can be further simplified for highly cited papers ($c \gg f_1$) as

$$\Delta f_r(c) \approx \frac{c}{C_1}, \quad \text{when } c \gg f_1. \qquad (5)$$

Let us now return to $\Delta f(c)$ and make a few remarks.

(a) For $c > f_1$, the additional paper is above-average with respect to the journal, and there is a *benefit* to publication: $\Delta f(c) > 0$ and the IF increases (i.e., $f_2 > f_1$).
(b) For $c < f_1$, the new paper is below-average with respect to the journal, and publishing it invokes a *penalty*: $\Delta f(c) < 0$ and the IF drops (i.e., $f_2 < f_1$).

(c) For $c = f_1$, the new paper is average, and publishing it makes no difference in the IF.
(d) Most important: The presence of $N_1$ in the denominator means that the benefit or penalty of publishing an additional paper decays rapidly with journal size. This has dramatic consequences, as we will see.

Let us now consider two special cases of interest:

> *Case 1.* The new paper is well above average relative to the journal, i.e., $c \gg f_1$. Here,

$$\Delta f(c) = \frac{c - f_1}{N_1 + 1} \approx \frac{c}{N_1 + 1} \approx \frac{c}{N_1}, \qquad (6)$$

where the last step is justified since in realistic cases we have $N_1 \gg 1$. The benefit $\Delta f(c)$ depends on the paper itself and on the journal size. As mentioned above, the presence of $N_1$ in the denominator means that publishing an above-average paper is far more beneficial to small journals than to large journals. For example, a journal *A* that is ten times smaller than a journal *B* will have a ten times higher benefit upon publishing the *same* highly cited paper, even if both journals had the same IF to begin with! The *editorial implication* here is that it pays for editors of small journals to be particularly watchful for high-performing papers. From the perspective of competing in IF rankings, small journals have two conflicting incentives: Be open to publishing risky and potentially breakthrough papers on the one hand, but not publish too many papers lest they lose their competitive advantage due to their small size.

For $c \ll N_1$, we have $\Delta f(c) \approx 0$, even when $c$ is large. This means that large journals, even when they publish highly cited papers, have a tiny benefit in their IF. For example, when a journal with $N_1 = 2000$ publishes a highly-cited paper of $c = 100$, the benefit is a mere $\Delta f(100) = 0.05$. For a very large journal of $N_1 = 20{,}000$, even an extremely highly cited paper of $c = 1000$ will produce a small gain of $\Delta f(1000) = 0.05$.

> *Case 2.* The new paper is well below average, i.e., $c \ll f_1$. Again, by "average" we mean with respect to the journal, not the global population of papers. (For journals of low IF, say, $f_1 \lesssim 2$, the condition $c \ll f_1$ implies $c = 0$.) Here,

$$\Delta f(c) = \frac{c - f_1}{N_1 + 1} \approx -\frac{f_1}{N_1 + 1} \approx -\frac{f_1}{N_1}, \qquad (7)$$

since in realistic cases we have $N_1 \gg 1$. The penalty $\Delta f(c)$ depends now only on the journal parameters $(N_1, f_1)$, and is greater for small-sized, high-IF journals. The *editorial implication* is that editors of small journals—and especially editors of small *and* high-IF journals—need to be more vigilant in pruning low-performing papers than editors of large journals. Two kinds of papers are low-cited, at least in the IF citation window: (a) archival, incremental papers, and (b) some truly ground-breaking papers that may appear too speculative at the time and take more than a couple years to be recognized.

For $f_1 \ll N_1$, we have $\Delta f(c) \approx 0$. So, very large journals have little to lose by publishing low-cited papers.

The take-home message from the above analysis is two-fold. First, with respect to increasing their IF, it pays for all journals take risks. Because the maximum penalty for publishing below-average papers ($\approx f_1/N_1$) is smaller in magnitude than the maximum benefit for publishing above-average papers ($\approx c/N_1$), it is better for a journal's IF that its editors publish a paper they are on the fence about, if what is at stake is the possibility of a highly influential paper that, if proven to be correct, may be ground-breaking. Some of these papers may also reap high citations to be worth the risk: recall that $c$ can lie in the hundreds or even thousands.

However, the reward for publishing breakthrough papers is much higher for small journals. For a journal's IF to seriously benefit from ground-breaking papers, the journal must above all remain small, otherwise the benefit is much reduced due to its inverse dependence with size. To the extent that editors of elite journals are influenced by IF considerations, they have an incentive to keep a tight lid on their risk-taking decisions and perhaps reject some potentially breakthrough research they might otherwise have published. We wonder whether the abundance of prestigious high-IF journals with biennial sizes smaller than $N_{2Y} < 400$ bears any connection to this realization.

On a related note, Wang, Veugelers and Stephan (2017) have reported on the increased difficulty of transformative papers to appear in prestigious journals. They found that "novel papers are less likely to be top cited when using short time-windows," and "are published in journals with Impact Factors lower than their non-novel counterparts, *ceteris paribus*." They argue that the increased pressure on journals to boost their IF "suggests that journals may strategically choose to not publish novel papers which are less likely to be highly cited in the short run." Our analysis may suggest an additional explanation for their findings that "novel papers encounter obstacles in being accepted by journals holding central positions in science" namely, the punishing effect of journal size on the IF.

**Systematic study of the volatility index $\Delta f(c)$, using data from 11,639 journals.**

Now let us look at some actual IF data. We ask the question: *How did the IF (citation average) of each journal change by incorporation of its most cited paper, which was cited $c^*$ times in the IF 2-year time-window?* We thus calculate the quantity $\Delta f(c^*)$, where $c^*$ is no longer constant and set equal to some theoretical value, but varies across journals.

First, some slight change in terminology to avoid confusion. We wish to study the effect of a journal's top-cited paper on its IF. Suppose the journal has a citation average $f$ and a biennial publication count $N_{2Y}$. So, our journal's "initial" state has size $N_1 = N_{2Y} - 1$ and citation average $f_1$, which we denote as $f^*$. Our journal's "final" state has $N_2 = N_{2Y}$ and $f_2 = f$, and was produced by incorporation of the top-cited paper that was cited $c^*$ times. We study how $\Delta f(c^*)$ and $\Delta f_r(c^*)$ behave using data from journals listed in the 2017 JCR.

Among the 12,266 journals initially listed in the 2017 JCR, we removed the several hundred duplicate entries, as well as the few journals whose IF was listed as zero or not available. We thus ended up with a master list of 11639 unique journal titles that received a 2017 IF as of December 2018. For each journal in this master list we obtained its Journal Citation Report, which contained the 2017 citations to each of its citable papers (i.e., articles and reviews) published in 2015–2016. We were thus able to calculate the citation average $f$ for each journal, namely, the ratio of 2017 citations to 2015–2016 citable papers. The citation average $f$ approximates the IF and becomes identical to it provided there are no "free" citations in the

numerator—that is, citations to *front-matter* items such as editorials, letters to the editor, commentaries, etc., or just "stray" citations to the journal without specific reference of volume and page or article number. We will thus use the terms "IF" and "citation average" interchangeably, for simplicity. Together, the 11639 journals in our master list published 3,088,511 papers in 2015–2016, which received 9,031,575 citations in 2017 according to the JCR. This is our data set.

In Fig. 1 we plot the volatility $\Delta f(c^*)$ vs. $N_{2Y}$ for each journal in our data set. In Table 1 we identify the top-10 journals in terms of $\Delta f(c^*)$, while in Table 2 we show the frequency distribution of $\Delta f(c^*)$ values. Finally, Tables 3 and 4 pertain to the relative volatility $\Delta f_r(c^*)$.

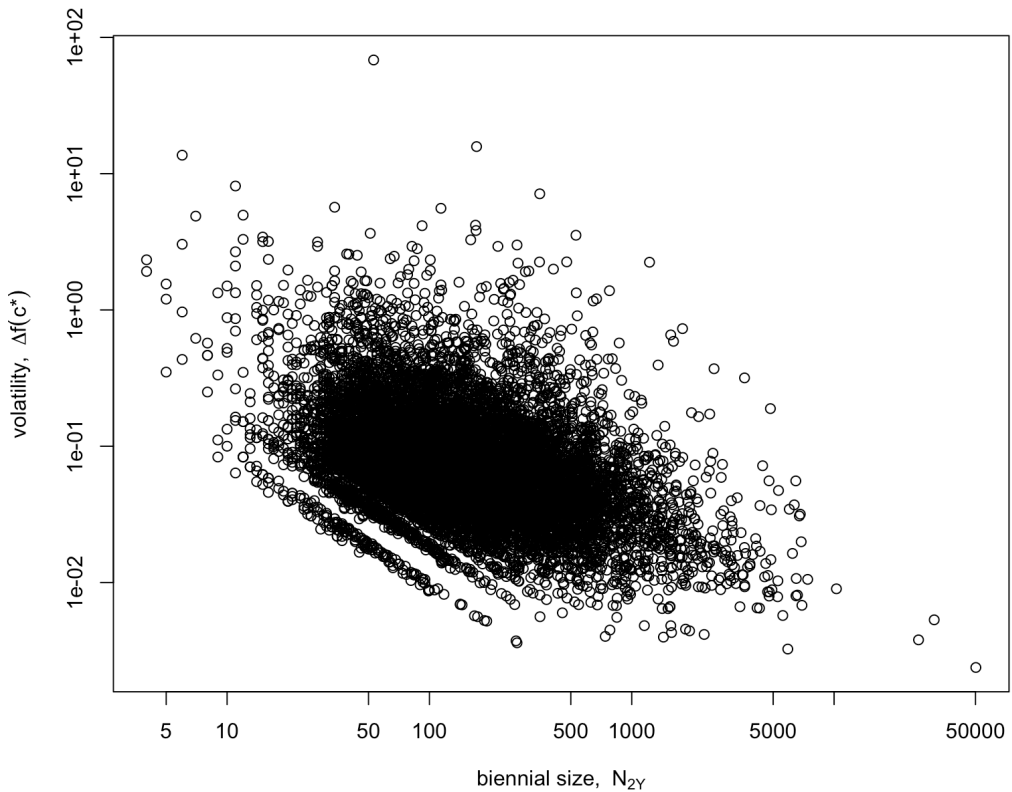

**Figure 1. IF volatility, $\Delta f(c^*)$, vs. journal biennial size, $N_{2Y}$, for all 11639 journals that received an IF in the 2017 JCR.**

Our key findings are as follows. A more detailed analysis will be presented in a forthcoming publication (Antonoyiannakis (2019, in preparation)).

1. *Large values of IF volatility occur for small journal sizes, namely, for $N_{2Y} \leq 2000$ and especially for $N_{2Y} \leq 500$.* (That is, for journals publishing *annually* less than 1000 and 250 citable items, respectively.) By large values of volatility, we mean $\Delta f(c^*) \approx 0.5$ and $\Delta f_r(c^*) \approx 25\%$, say.

2. *Many journals experience a large boost in their IF due to their most cited paper.* For instance (see Table 2), there are 381 journals in our data set where $\Delta f(c^*) > 0.5$, i.e., a single paper raises a journal's citation average by at least half a point. For 140 journals we have $\Delta f(c^*) > 1$, while for 41 journals we have $\Delta f(c^*) > 2$, and so on.

3. *For some journals, an extremely highly cited paper causes a large $\Delta f(c^*)$ value.* Consider the top 2 journals in Table 1. The journal *CA-A Cancer Journal for Clinicians* published in 2016 a research article that was cited 3790 times in 2017, which accounted for almost 30% of the total citations that entered in its IF calculation that year, with a corresponding $\Delta f(c^*) = 68.3$. Without this paper, the journal's citation average would have dropped from $f = 240.1$ to a "meager" $f^* = 171.8$. Similarly, the *Journal of Statistical Software* published in 2015 a research article that gathered 2708 citations in 2017 and captured 73% of the total citations to the journal that year. Although such extreme levels of volatility are rare, they do occur every year, because of papers cited thousands of times and published in small journals.

4. *A paper needs not be exceptionally cited to produce a large IF boost provided the journal is sufficiently small.* Consider the journals in positions #3 and #4 in Table 1, namely, *Living Reviews in Relativity* and *Psychological Inquiry*. These journals' IFs were strongly boosted by their top-cited paper, even though the latter was much less cited ($c^*=87$ and $c^*=97$, respectively) than for the top 2 journals. This happened because journal sizes were smaller also ($N_{2Y} = 6$ and 11, respectively). Such occurrences are not uncommon, because papers cited dozens of times are much more abundant than papers cited thousands of times, while there are also plenty of very small journals. Indeed, within the top-40 journals (not shown here) in terms of decreasing volatility there are 3 journals whose top-cited paper received 32, 42, and 13 citations respectively, causing a significant $\Delta f(c^*)$ that ranged from 2.3 to 2.6. High values of *relative* volatility $\Delta f_r(c^*)$ due to low-cited or moderately-cited papers are much more common—see Table 3 and journals in positions #2, #3, #4, #9, and #10.

Table 1. **Top-10 journals in volatility $\Delta f(c^*)$, i.e., *absolute* change in IF due to their top-cited paper.**

|    | Journal              | $\Delta f(c^*)$ | $c^*$ | $\Delta f_r(c^*)$ | $f$    | $f^*$  | $N_{2Y}$ |
|----|----------------------|-----------------|-------|-------------------|--------|--------|----------|
| 1  | CA-CANCER J CLIN     | **68.27**       | 3790  | 40 %              | 240.09 | 171.83 | 53       |
| 2  | J STAT SOFTW         | **15.80**       | 2708  | 271%              | 21.63  | 5.82   | 171      |
| 3  | LIVING REV RELATIV   | **13.67**       | 87    | 273%              | 18.67  | 5.00   | 6        |
| 4  | PSYCHOL INQ          | **8.12**        | 97    | 105%              | 15.82  | 7.70   | 11       |
| 5  | ACTA CRYSTALLOGR C   | **7.12**        | 2499  | 474%              | 8.62   | 1.50   | 351      |
| 6  | ANNU REV CONDEN MA P | **5.67**        | 209   | 35%               | 21.82  | 16.15  | 34       |
| 7  | ACTA CRYSTALLOGR A   | **5.57**        | 637   | 271%              | 7.62   | 2.05   | 114      |
| 8  | ADV PHYS             | **4.96**        | 85    | 19%               | 30.42  | 25.45  | 12       |
| 9  | PSYCHOL SCI PUBL INT | **4.88**        | 49    | 33%               | 19.71  | 14.83  | 7        |
| 10 | ACTA CRYSTALLOGR B   | **4.19**        | 710   | 199%              | 6.30   | 2.11   | 169      |

Table 2. **Number of journals whose volatility $\Delta f(c^*)$ was *greater than* the threshold value listed in the 1st column.**

| $\Delta f(c^*)$ | No. journals above threshold | % all journals |
|-----------------|------------------------------|----------------|
| 0.1             | 3881                         | 33.3%          |
| 0.25            | 1061                         | 9.1%           |
| 0.5             | 381                          | 3.3%           |
| 0.75            | 221                          | 1.9%           |

| 1   | 140 | 1.2%  |
|-----|-----|-------|
| 1.5 | 73  | 0.63% |
| 2   | 41  | 0.35% |
| 3   | 21  | 0.18% |
| 4   | 11  | 0.09% |
| 5   | 7   | 0.06% |
| 10  | 3   | 0.03% |
| 50  | 1   | 0.01% |

**Table 3**. Top-10 journals in *relative* volatility $\Delta f_r(c^*)$, i.e., relative change in IF due to their top-cited paper.

|    | Journal           | $\Delta f(c^*)$ | $c^*$ | $\Delta f_r(c^*)$ | $f$   | $f^*$ | $N_{2Y}$ |
|----|-------------------|-----------------|-------|-------------------|-------|-------|----------|
| 1  | ACTA CRYSTALLOGR C| 7.12            | 2499  | **474%**          | 8.62  | 1.50  | 351      |
| 2  | COMPUT AIDED SURG | 0.88            | 9     | **395%**          | 1.10  | 0.22  | 10       |
| 3  | ETIKK PRAKSIS     | 0.15            | 4     | **381%**          | 0.19  | 0.04  | 26       |
| 4  | SOLID STATE PHYS  | 3.03            | 19    | **379%**          | 3.83  | 0.80  | 6        |
| 5  | CHINESE PHYS C    | 2.25            | 1075  | **350%**          | 2.90  | 0.64  | 477      |
| 6  | LIVING REV RELATIV| 13.67           | 87    | **273%**          | 18.67 | 5.00  | 6        |
| 7  | J STAT SOFTW      | 15.80           | 2708  | **271%**          | 21.63 | 5.82  | 171      |
| 8  | ACTA CRYSTALLOGR A| 5.57            | 637   | **271%**          | 7.62  | 2.05  | 114      |
| 9  | AFR LINGUIST      | 0.26            | 3     | **264%**          | 0.36  | 0.10  | 11       |
| 10 | AM LAB            | 0.04            | 5     | **247%**          | 0.05  | 0.01  | 136      |

**Table 4.** Number of journals whose *relative* volatility $\Delta f_r(c^*)$ was *greater than* the threshold value listed in the 1st column.

| $\Delta f_r(c^*)$ | No. journals above threshold | % all journals |
|-------------------|------------------------------|----------------|
| 10%               | 3403                         | 29.2%          |
| 20%               | 1218                         | 10.5%          |
| 25%               | 818                          | 7%             |
| 30%               | 592                          | 5.1%           |
| 40%               | 387                          | 3.3%           |
| 50%               | 231                          | 2.0%           |
| 60%               | 174                          | 1.5%           |
| 70%               | 140                          | 1.2%           |
| 80%               | 124                          | 1.07%          |
| 90%               | 114                          | 0.87%          |
| 100%              | 50                           | 0.43%          |
| 300%              | 5                            | 0.04%          |

**Conclusions**

The above findings corroborate our earlier conclusion (Antonoyiannakis, 2018) that IFs are scale dependent and particularly volatile to small journal sizes, as explained by the Central Limit Theorem. This point is pertinent for real journals because 90% of all journals publish no more than 250 citable items annually (Antonoyiannakis, 2018).

Compared to large journals, small journals have (a) much more to gain by publishing a highly-cited paper, and (b) more to lose by publishing a little-cited paper—that is, more to gain by rejecting a little-cited paper. Therefore, in terms of IF, it pays for small journals to be selective.

The fact that there are more than a hundred journals annually whose highest cited paper suffices to raise their citation average by 1 point demonstrates that the effect we study here is not of academic but of practical interest. If so many journals are that much affected by a single paper, then the usefulness of the IF as a *journal* defining quantity is questioned, as is the practice of IF rankings of journals. This point becomes even more pertinent when we consider the *relative* volatility. Evidently (see Table 4), for 1 out of 50 journals (231 journals) a single paper boosts the IF by 50%. Roughly 1 out of 10 journals (1218 journals) had their IF boosted by more than 20% by a single paper. And for more than a quarter of all journals (3403 journals) the IF increased more than 10% by a single paper.

So, the IF volatility affects thousands of journals. It is not an exclusive feature of a few journals or a statistical anomaly that we can casually brush off, but an everyday feature that is inherent in citation averages (IFs) and affects many journals, every year.

The high volatility of IF values from real-journal data demonstrates that ranking journals by IFs constitutes a non-level playing field, since the IF gain of publishing an equally cited paper scales as the inverse journal size and can therefore span up to 4 orders of magnitude across journals. It is therefore critical to consider novel ways of comparing journals based on more solid statistical grounds. The implications of such a decision may reach much further than producing ranked journal lists aimed at librarians (the original motivation for the IF) and affect research assessment and the careers or scientists.

*Disclaimer*: The author is an Associate Editor at the American Physical Society. These opinions are his own.